\documentclass[11pt,tightenlines,aps,prd,groupedaddress,nofootinbib,longbibliography,notitlepage]{revtex4-1}

\usepackage{amsmath,amssymb,bm,mathtools}

\usepackage{graphicx}
\usepackage{amsfonts}
\usepackage{amssymb}
\usepackage{amsbsy}
\usepackage{amsmath}
\usepackage{latexsym}
\usepackage{bm}
\usepackage{color}
\usepackage{hyperref}
\usepackage{comment}
\usepackage[normalem]{ulem}
\usepackage[makeroom]{cancel}

\usepackage{stackengine}
\stackMath
\newcommand\tenq[2][1]{%
 \def\useanchorwidth{T}%
  \ifnum#1>1%
    \stackon[0pt]{\tenq[\numexpr#1-1\relax]{#2}}{\scriptscriptstyle\sim}%
  \else%
    \stackon[1pt]{#2}{\scriptscriptstyle\sim}%
  \fi%
}

\newcommand{\be}{\begin{equation}}
\newcommand{\ee}{\end{equation}}
\newcommand{\de}{\mbox{d}}
\newcommand{\f}{\frac}

\newcommand{\pa}{\partial}
\newcommand{\pha}{\phantom{a}}
\newcommand{\scr}{\scriptscriptstyle}

\newcommand{\cmmnt}[1]{}

\numberwithin{equation}{section}

\begin{document}

\title{Interacting dark sector from the trace-free Einstein equations: cosmological perturbations with no instability}

\author{Marco de Cesare$^{*, \dagger}$ and Edward Wilson-Ewing$^{\dagger}$}

\affiliation{* Department of Physics, University of the Basque Country UPV/EHU, 48940 Leioa, Spain \\
$\dagger$ Department of Mathematics and Statistics, University of New Brunswick, Fredericton, NB, Canada E3B 5A3}

\email{marco.decesare@ehu.eus}
\email{edward.wilson-ewing@unb.ca}

\begin{abstract}
In trace-free Einstein gravity, the stress-energy tensor of matter is not necessarily conserved and so the theory offers a natural framework for interacting dark energy models where dark energy has a constant equation of state $w=-1$.  We derive the equations of motion for linear cosmological perturbations in interacting dark energy models of this class, focusing on the scalar sector.  Then, we consider a specific model where the energy-momentum transfer potential is proportional to the energy density of cold dark matter; this transfer potential has the effect of inducing an effective equation of state $w_{\rm eff}\neq0$ for cold dark matter. We analyze in detail the evolution of perturbations during radiation domination on super-Hubble scales, finding that the well-known large-scale instability that affects a large class of interacting dark energy models is absent in this model. To avoid a gradient instability, energy must flow from dark matter to dark energy. Finally, we show that interacting dark energy models with $w=-1$ are equivalent to a class of generalized dark matter models.
\end{abstract}

\maketitle

\section{Introduction}

Trace-free Einstein gravity is a minimal modification of general relativity where the Einstein field equations are replaced by their trace-free counterpart~\cite{Ellis:2010uc}. There is one less dynamical equation and as a consequence the evolution of the conformal degree of freedom in the metric cannot be fully determined without making further assumptions. In fact, the covariant conservation law for the stress-energy tensor of matter, which in general relativity follows from the contracted Bianchi identities, here needs to be imposed by hand \cite{Ellis:2010uc}. If we refrain from making this extra assumption, then the trace-free Einstein equations can be recast in the form of the usual Einstein equations except with a space-time dependent cosmological ``constant'' $\Lambda(x^a)$, whose evolution is determined by the exchange of energy-momentum with matter \cite{Josset:2016vrq}.  This naturally provides a class of interacting dark energy (IDE) models which are embedded within the gravitational theory based on the trace-free Einstein equations.

Interacting dark energy models have been widely studied as phenomenologically motivated extensions of standard cosmology. The interactions with dark energy may be restricted to the dark sector, or may extend to baryons. Originally proposed in order to alleviate the cosmic coincidence problem \cite{Wang:2016lxa}, IDE models also affect structure growth~\cite{Clemson:2011an,Caldera-Cabral:2009hoy,Amendola:2001rc,Chamings:2019kcl}, and have recently gained new interest as a possible solution to the Hubble tension, see, e.g., \cite{DiValentino:2021izs} and the many references therein. However, many IDE models are affected by a large-scale instability, which poses severe restrictions on their viability \cite{Valiviita:2008iv}. In particular, the instability is present in some widely studied phenomenological models, where dark energy has a constant equation-of-state parameter $w_{\scriptscriptstyle \rm DE}$; specific examples include models where the background energy transfer is proportional to $\rho_{\scriptscriptstyle \rm CDM}$\,, $\mathcal{H} \rho_{\scriptscriptstyle \rm CDM}$\,, or $\mathcal{H} (\rho_{\scriptscriptstyle \rm CDM} + \rho_{\scriptscriptstyle \rm DE})$ \cite{Valiviita:2008iv}. One way to cure the instability is by allowing for a time-dependent equation of state, $w_{\scriptscriptstyle \rm DE}^{\prime}\neq0$ \cite{Majerotto:2009np}, and a model with background energy transfer proportional to the dark energy density $\rho_{\scriptscriptstyle \rm DE}$ was also shown to be free from large-scale instabilities \cite{Clemson:2011an}. In the context of general relativity, IDE models with $w=-1$ (interacting vacuum) have been studied in Ref.~\cite{Wands:2012vg} and a phase space analysis was carried out in Ref.~\cite{Kaeonikhom:2020fqs}; a more general treatment based on the parametrized post Friedmann (PPF) formalism has constrained the parameter space for a large class of IDE models \cite{Skordis:2015yra}, while observational constraints for some IDE models have also been obtained \cite{Valiviita:2009nu, Clemson:2011an, Guo:2017hea, DiValentino:2017zyq, vonMarttens:2018iav, Martinelli:2019dau, Yang:2019uog}.

Even though IDE models are phenomenologically interesting, most models lack guidance from fundamental physics that could explain the origin and characteristics of interactions in the dark sector. One interesting possibility is that violations of energy-momentum conservation may originate from dissipative processes taking place at the Planck scale \cite{Perez:2017krv}. This can be reconciled with general covariance if the trace-free Einstein equations are assumed as a low-energy effective description of quantum gravity, where $\nabla^b T_{ab} \neq 0$ and this energy loss sources a dynamical $\Lambda$. Within this framework, the possibility of alleviating the Hubble tension has been studied \cite{Perez:2020cwa} and observational constraints for similar models have been obtained \cite{Corral:2020lxt, LinaresCedeno:2020uxx}. Furthermore, the possibility of an inflationary mechanism based on the relaxation of the cosmological constant has been investigated \cite{Amadei:2021aqd,Leon:2022kwn}.

In this paper, we give a general analysis of scalar cosmological perturbations using the trace-free Einstein equations, assuming an energy-momentum transfer between ordinary matter and dark energy, and then specialize to a model where the energy-momentum transfer only takes place between dark matter and dark energy, and the transfer potential is proportional to the energy density of cold dark matter. An important result is that in this model there is no instability for linear perturbations; this is an important difference with many other IDE models including the ones examined in Ref.~\cite{Valiviita:2008iv}. As explained in Sec.~\ref{Sec:Dsummary}, the usual instability is driven by the non-adiabatic pressure perturbation of the dark energy fluid, but for the trace-free Einstein equations and a transfer potential that depends only on the dark matter energy density there is no non-adiabatic pressure perturbation to dark energy---this is ultimately why there is no instability here.

Note that the trace-free Einstein equations are closely linked to unimodular gravity, since the equations of motion of unimodular gravity are precisely the trace-free Einstein equations. However, in unimodular gravity there is also an additional unimodularity constraint that breaks the full diffeomorphism invariance of general relativity down to volume-preserving diffeomorphisms; this unimodularity constraint is introduced so the trace-free Einstein equations can be derived from an action principle \cite{Percacci:2017fsy}. (For other actions that give the trace-free equations of unimodular gravity, see Refs.~\cite{Henneaux:1989zc, Kuchar:1991xd, Jirousek:2018ago}). The quantum theory of unimodular gravity has been studied in Refs.~\cite{deLeonArdon:2017qzg, deBrito:2021pmw}.

Cosmological perturbations in unimodular gravity have been studied previously \cite{Gao:2014nia, Basak:2015swx}, but since in these earlier studies the space-time metric is also subject to the unimodularity constraint (that breaks the full diffeomorphism invariance of general relativity), as a consequence some convenient gauges that are commonly used in general relativity, like the longitudinal and the synchronous gauges, are not available. In our approach, we do not assume an action principle, and instead take as a starting point only the trace-free Einstein equations, which retain the full diffeomorphism invariance of general relativity (and also of standard cosmological perturbation theory). Another important difference lies in the fact that we do not impose the conservation of the stress-energy tensor of ordinary matter: this violation naturally leads to interacting dark energy as already explained.  Yet another approach to the cosmology of trace-free Einstein gravity has also been proposed where the condition that the energy-momentum transfer to/from dark energy be exact is not imposed \cite{Garcia-Aspeitia:2019yni, Garcia-Aspeitia:2019yod} (for more on this condition, see Sec.~\ref{Sec:FieldEqs_General}); this, combined with the Bianchi identities, gives a third-order equation for the total stress-energy of matter. In contrast, the equations of motion for perturbations derived here are all second-order.

The paper is organized as follows. In Sec.~\ref{Sec:FieldEqs_General}, we review the relation between the trace-free Einstein equations and IDE models with integrable energy-momentum transfer, and recast the gravitational field equations as effective Einstein field equations with a space-time-dependent $\Lambda$. In Sec.~\ref{Sec:Framework}, using standard techniques we derive the equations for the cosmological background and (linear) perturbations for the trace-free Einstein equations, and the comparison with some other transfer models studied earlier in Ref.~\cite{Valiviita:2008iv} is presented in Sec.~\ref{Sec:Comparison}. In Sec.~\ref{Sec:TransferModel}, we introduce a specific model for the covariant energy-momentum transfer, from which we derive the resulting equations of motion for the background and perturbations. Our main results are in Secs.~\ref{Sec:RDE} and \ref{Sec:GMDE}, where we solve for the dynamics of cosmological perturbations during the radiation dominated era and during dark matter domination, respectively. We discuss possible generalizations of the model in Sec.~\ref{Sec:Generalizations}, and in Sec.~\ref{Sec:ConnectionGDM} we show that this family of IDEs is equivalent to a class of generalized dark matter models \cite{Hu:1998kj}. Finally, we present our conclusions in Sec.~\ref{Sec:Conclusions}.

\bigskip

\noindent
{\bf Conventions:} In our units, $c=1$ for the speed of light, and the gravitational coupling is denoted as $\kappa=8\pi G$. We assume the space-time signature $(- + + +)$. A prime $^\prime$ denotes a derivative w.r.t.~conformal time, and a comma is used to denote partial derivatives w.r.t.~comoving coordinates: $f_{,i}\equiv\pa_i f$. The position (raised or lowered) of labels denoting a given matter species is irrelevant. We will use the label $x$ to denote dark energy, $c$ to denote cold dark matter, $r$ to denote radiation, and $b$ to denote baryonic matter.

\section{Trace-free Einstein equations}
\label{Sec:FieldEqs_General}

We assume the dynamics of the gravitational field satisfy the trace-free Einstein equations
\be\label{Eq:TracelessFieldEquations}
R_{ab}-\frac{1}{4}R\, g_{ab}=\kappa \left(T_{ab}- \frac{1}{4}T g_{ab}\right)~,
\ee
where $T_{ab}$ represents the stress-energy tensor of matter. Note that, unlike general relativity, the stress-energy of matter is not necessarily conserved \cite{Josset:2016vrq, Perez:2017krv}. In fact, taking the divergence of both sides of Eq.~\eqref{Eq:TracelessFieldEquations} and using the contracted Bianchi identities gives
\be\label{Eq:NonConservation}
\kappa \nabla^{c}T_{ac}=\frac{1}{4}\nabla_a(R+\kappa T)\eqqcolon J_a ~.
\ee
$J_a$ is the energy-momentum transfer, which measures the extent to which stress-energy conservation is violated.
We stress that we only assume the field equations~\eqref{Eq:TracelessFieldEquations} without imposing any further constraints, such as the unimodularity condition. We discuss this point further at the end of this subsection.

The trace-free field equations can be rewritten as
\be 
R_{ab} - \f{1}{2} R \, g_{ab} + \f{1}{4} (R + \kappa T) g_{ab} = \kappa T_{ab},
\ee
and then, using \eqref{Eq:NonConservation} gives
\be \label{Eq:Integ}
R_{ab} - \f{1}{2} R \, g_{ab} + \left( \Lambda_{\rm f} + \int_\ell J \right) g_{ab} = \kappa T_{ab},
\ee
where $\Lambda_{\rm f}$ is an integration constant and $\ell$ is any path from some fixed space-time point to $x$. With an appropriate choice of the path $\ell$,  $\Lambda_{\rm f}$ can be interpreted as the value of the cosmological constant in the distant future. To ensure that this equation is independent of the path $\ell$ and only depends on its endpoint $x$, we only allow the case where $J_a$ obeys the integrability condition $\de J=0$ \cite{Josset:2016vrq, Perez:2017krv}, which locally implies%
\footnote{The integrability condition $\de J=0$ was initially derived by assuming invariance under volume-preserving diffeomorphisms \cite{Josset:2016vrq}. If instead one only assumes the field equations \eqref{Eq:TracelessFieldEquations} as a starting point, the integrability condition remains necessary to ensure that the Einstein tensor be single-valued as can be seen in Eq.~\eqref{Eq:Integ}.} 
\be\label{Eq:DefQ}
J_{a}=-\nabla_a Q~,
\ee
for some scalar function $Q=Q(x)$, which is the energy-momentum transfer potential. (For a discussion on a possible microscopic origin of energy-momentum non-conservation, see~\cite{Josset:2016vrq, Perez:2017krv}.) With Eq.~\eqref{Eq:DefQ}, the field equations~\eqref{Eq:TracelessFieldEquations} can be rewritten as
\be\label{Eq:EffectiveEFE}
G_{ab}=\kappa (T_{ab}+\tilde{T}_{ab})~,
\ee
where the effective stress-energy tensor is
\be\label{Eq:EffectiveDE}
\kappa\, \tilde{T}_{ab}=Q\, g_{ab}~.
\ee
Note that $\nabla^a(T_{ab} + \tilde T_{ab}) = 0$ by construction. The effective stress-energy tensor $\tilde{T}_{ab}$ can be interpreted as corresponding to a space-time-dependent cosmological ``constant'' $\Lambda_{\rm eff}(x)=-Q(x)$~, whose variation in space and time is determined by the non-conservation of the stress-energy of ordinary matter \cite{Josset:2016vrq, Perez:2017krv}. Thus, the field equations~\eqref{Eq:EffectiveEFE} describe an interacting dark energy model.

If there are several matter components (each labelled by an index $A$), the total matter stress-energy is
\be\label{Eq:MatterTab}
T_{ab}=\sum_A T_{ab}^{A}~,
\ee
and for each term contributing to Eq.~\eqref{Eq:MatterTab},
\be\label{Eq:ContinuityFullA}
\kappa \nabla^b T_{ab}^{A}=J_a^A~,
\ee
where the energy-momentum transfer vectors must satisfy
\be\label{Eq:BalanceJ}
\sum_A J_a^A=J_a ~.
\ee
In general, the individual contributions $J_c^A$ include terms corresponding to energy-momentum transfer between different matter species, as well as to dark energy: here we will focus on the second effect. We will assume that all fluxes $J_c^A$ are integrable:
\be\label{Eq:DefJ_A}
J_a^A=-\nabla_a Q^A, \qquad \qquad
\sum_A Q^A= Q ~.
\ee
It is also convenient to define the energy density of the effective dark energy fluid as
\be\label{Eq:DefRhoX}
\rho_x=-\kappa^{-1}Q ~,
\ee
and by combining Eqs.~\eqref{Eq:NonConservation},~\eqref{Eq:EffectiveDE} and \eqref{Eq:DefRhoX}, the continuity equation for $T_{ab}+\tilde{T}_{ab}$ can be rewritten as
\be\label{Eq:Continuity}
\nabla^{b}T_{ab}=\nabla_a \rho_x~.
\ee
Note that even though $\rho_x$ is not constant, the equation of state $w_x=-1$ for the effective dark energy fluid is exact.

\bigskip

\noindent
\textbf{Remarks on the unimodularity condition:} \\
In order to derive the trace-free Einstein equations \eqref{Eq:TracelessFieldEquations} from an action principle, one can add to the Einstein-Hilbert action $\tfrac{1}{2\kappa}\int \de^4x\sqrt{-g}\, R$ a new term $\int \de^4x\, \lambda\left( \omega-\sqrt{-g}\,\right)$, where $\omega$ is a fixed non-dynamical volume form, and $\lambda$ is a Lagrange multiplier \cite{Percacci:2017fsy}. This new term enforces the so-called unimodularity condition $\sqrt{-g}=\omega$~. The unimodularity condition can be regarded as a gauge-fixing condition, which removes one local (gauge) degree of freedom and fixes the (global) scale factor. The local physical degrees of freedom are the same as in general relativity, but the residual gauge symmetries are now given by the group of volume-preserving diffeomorphisms.

This action principle is assumed as a starting point in some previous studies of cosmological perturbations in unimodular gravity \cite{Gao:2014nia, Basak:2015swx}, in which case the perturbed trace-free equations are supplemented by the condition $\sqrt{-g}=\omega$, perturbed to first order, which clearly restricts the class of allowed gauges, see also the discussion in Ref.~\cite{Fabris:2021atr}.  (Using the notation of Sec.~\ref{Sec:Framework}, to first order in perturbation theory the unimodularity condition gives $\nabla^2 E +\phi- 3\psi=0$.)

In our approach, we do not use this action principle (an action principle is in tension with the idea that the interacting dark energy is due to coarse-grained diffusion effects \cite{Perez:2017krv}); instead our starting point is the trace-free Einstein equations only.  As a result, we do not assume the condition~$\sqrt{-g}=\omega$ or $\nabla^2 E +\phi- 3\psi=0$.

\section{Cosmological background and scalar perturbations}
\label{Sec:Framework}

In this section we derive the cosmological evolution equations for both the background and first-order perturbations, taking Eq.~\eqref{Eq:EffectiveEFE} as a starting point. Since the effective stress-energy in Eq.~\eqref{Eq:EffectiveEFE} is completely specified by the scalar function $Q$, it has no effects on the dynamics of vector and tensor perturbations to linear order in perturbation theory; therefore, we will only study its effects on the evolution of scalar perturbations. Given the similarities between our model and the interacting dark sector models studied in Ref.~\cite{Valiviita:2008iv}, our notation and analysis will be similar. The equations derived in this section are fully general and do not rely on a specific choice of the potential $Q$.

Assuming a spatially flat FLRW background, the perturbed line element is \cite{Mukhanov:1990me}
\be\label{Eq:perturbed_metric}
\de s^2=a(\eta)^2 \left\{-(1+2\phi)\de\eta^2+2B_{,i}\de x^i\de\eta+ \left[(1-2\psi)\delta_{ij}+2E_{,ij}\right] \de x^i\de x^j  \right\} ~.
\ee
Each matter species (labelled by the index $A$) is assumed to be hydrodynamic matter with the stress-energy tensor
\be
T_{ab}^A=(\rho_A+p_A) u_a^A u_b^A+ p_A\, g_{ab}+\pi^A_{ab} ~.
\ee
For each component, we assume a barotropic equation of state $p_A=w_A\, \rho_A$, where the $w_A$ are not necessarily constant. The four-velocities $u_a^A$ specify the rest-frame of each fluid and are normalized, $u_a^A u^{A\, a}=-1$~, while the anisotropic stress satisfies the two conditions $\pi^A_{ab} u^{A\, a}=0$~,~~$g^{ab}\pi^A_{ab}=0$.

When combining the contributions of different matter fields to form the total matter stress-energy $T_{ab}=\sum_A T_{ab}^A$~, the total four-velocity is not uniquely specified. We choose to define the total four-velocity $u^a$ as a time-like eigenvector of $T_{ab}$, in this way the net momentum flux is zero in the rest frame specified by $u^a$ \cite{Landau:1987}. This gives
\be\label{Eq:TabLandauFrame}
T_{ab} = (\rho+p) u_a u_b+ p\, g_{ab} + \pi_{ab} = \sum_A \left[ (\rho_A+p_A) u_a^A u_b^A + p_A \, g_{ab} + \pi^A_{ab} \right] ~.
\ee

\subsection{Background evolution of flat FLRW}
At the background level, spatial isotropy requires that the four-velocities of all fluids must be the same $u_a^A=\bar{u}_a$ for all $A$, and their anisotropic stress must be zero $\bar{\pi}^A_{ab}=0$. Normalizing the background four-velocity gives
\be
\bar{u}_a=-a(1,0)~,~~\bar{u}^a=a^{-1}(1,0)~,
\ee
so the background stress-energy tensor of each matter species is
\be
\overline{T}^{a}_{A\; b}=(\bar{\rho}_A+\bar{p}_A) \bar{u}^a \bar{u}_b+ \bar{p}_A\, \delta^a_{\pha b} ~,
\ee
with components
\be
\overline{T}^{0}_{A\; 0}=-\bar{\rho}_A~,~~\overline{T}^{i}_{A\; j}=\bar{p}_A\,\delta^i_{\; j}~.
\ee
For the effective stress-energy tensor $\tilde T_{ab}$ on the right side of Eq.~\eqref{Eq:EffectiveEFE},
\be
\overline{\tilde{T}}^{0}_{\pha 0}=-\bar{\rho}_x ~,~~\overline{\tilde{T}}^{i}_{\; j}=\bar{p}_x=-\bar{\rho}_x~.
\ee
The background evolution equations then read
\begin{subequations}
\begin{align}
\mathcal{H}^2&=\frac{\kappa }{3}a^2\left(\sum_A\bar{\rho}_A+\bar{\rho}_x\right)\eqqcolon \frac{\kappa }{3}a^2\left(\bar{\rho}+\bar{\rho}_x\right)  ~, \label{Eq:Fried1} \\
\mathcal{H}^2-2\frac{a^{\prime\prime}}{a}&=\kappa\, a^2\left(\sum_A\bar{p}_A-\bar{\rho}_x\right)\eqqcolon \kappa\, a^2\left(\bar{p}-\bar{\rho}_x\right) ~, \label{Eq:Fried2}
\end{align}
\end{subequations}
where $\mathcal{H}\coloneqq a^{\prime}/a$~.

Projecting the continuity equation \eqref{Eq:ContinuityFullA} for the matter species $A$ onto $\bar{u}^a$ gives
\be
\bar{\rho}_A^{\prime}+3\mathcal{H}(\bar{\rho}_A+\bar{p}_A)=-\kappa^{-1}  a \left(\bar{u}^c\bar{J}^A_c\right)=\kappa^{-1}\bar{Q}^{\prime}_A~,
\ee
subject to the second relation in Eq.~\eqref{Eq:DefJ_A}. Similarly, the continuity equation for the total stress-energy~\eqref{Eq:Continuity} implies
\be\label{Eq:ContinuityBckgrnd}
\bar{\rho}^{\prime}+3\mathcal{H}(\bar{\rho}+\bar{p})=\kappa^{-1}Q^{\prime}=-\bar{\rho}^{\prime}_x~.
\ee
Equation~\eqref{Eq:ContinuityBckgrnd} shows that, with our conventions, there is a net transfer of energy from ordinary matter to dark energy for $Q^{\prime}<0$ (and vice versa, for $Q^{\prime}>0$ the transfer takes place in the opposite direction).

Note that in order to solve the background dynamics as described by Eqs.~\eqref{Eq:Fried1}, \eqref{Eq:Fried2}, \eqref{Eq:ContinuityBckgrnd}, it is necessary to also assume a concrete model for the transfer of energy between ordinary matter and the effective dark energy, since the evolution of $\rho_x$ cannot be determined from the field equations alone. Some models have been proposed in Ref.~\cite{Perez:2017krv, Perez:2020cwa}; we will return to this point in Sec.~\ref{Sec:TransferModel}. Until then, we will leave $Q$ free and keep our analysis completely general.

\subsection{Perturbed stress-energy of matter}
\label{Sec:Pert-T}

The perturbed four-velocity for the fluid $A$ is
\be\label{Eq:PerturbedVelocityA}
u^a_A=a^{-1}\left(1-\phi,v^i_A\right)~,
\ee
where $v^i_A$ is the peculiar velocity (with respect to comoving coordinates), and its covariant counterpart is
\be
u_{a}^{A}=g_{ab}u^b_A = -a \left(1+\phi, -v_{i}^{A} - B_{,i} \right)~.
\ee
Since here we are only interested in scalar perturbations, we assume that the velocity field is irrotational, i.e.,  $v_i^A=v_{,i}^A$~, where $v^A$ is the peculiar velocity potential.

The perturbed stress-energy tensor for each matter species has the form
\be
\delta T^{a}_{A\; b}=(\delta \rho_A+\delta p_A) \bar{u}^{a} \bar{u}_{b}+(\bar{\rho}_A+\bar{p}_A)\left( \delta u^{a}_A \bar{u}_{b}+ \bar{u}^{a} \delta u_{b}^{A}\right)+  \delta{p}_A\, \delta^a_{\pha b}+\pi^{a}_{A\; b}~,
\ee
and the components are
\begin{subequations}\label{Eq:PerturbedTmunu}
\begin{align}
\delta T^0_{A\; 0}&=-\delta \rho_A~,\\
\delta T^0_{A\; i}&=(\bar{\rho}_A+\bar{p}_A)(v_{i}^{A}+B_{i})~,\\
\delta T^i_{A\; 0}&=-(\bar{\rho}_A+\bar{p}_A)v^i_A~,\\
\delta T^i_{A\; j}&=\delta p_A\, \delta^i_{\;  j} +\pi^i_{A\; j}~.
\end{align}
\end{subequations}
The momentum density is defined as $q^i_A\coloneqq (\bar{\rho}_A+\bar{p}_A)v^i_A$~, and the scalar potential $\pi_A$ for the anisotropic stress is
\be
\pi^{i}_{A\; j}=\left(\pa^i\pa_j -\frac{1}{3}\delta^{i}_{\; j} \triangle\right) \pi_A~.
\ee
From this point on, in equations involving only three-dimensional tensors on spatial slices, Latin indices will be raised and lowered with $\delta_{ij}$. Note that our definition of $\pi_A$ differs from the one in Ref.~\cite{Valiviita:2008iv} by an overall factor of $a^2$.

Let us now introduce the perturbative expansion for the four-velocity of the total fluid defined in \eqref{Eq:TabLandauFrame}
\be\label{Eq:PerturbedVelocityTotal}
u^a=a^{-1}\left(1-\phi,v^i\right)~,
\ee
where $v$ is the corresponding velocity potential. Combining Eqs.~\eqref{Eq:PerturbedVelocityA},~\eqref{Eq:PerturbedVelocityTotal},~\eqref{Eq:TabLandauFrame}, the physical quantities that characterize the perturbed total stress-energy tensor can be expressed in terms of its component fluids,
\be\label{Eq:PerturbedFluid}
\delta \rho= \sum_A \delta \rho_A~,~~\delta p= \sum_A \delta p_A~,~~  \pi= \sum_A \pi_A~,~~ (\bar{\rho}+\bar{p}) v= \sum_A (\bar{\rho}_A+\bar{p}_A) v_A~.
\ee
The last equation in~\eqref{Eq:PerturbedFluid} provides an explicit definition for $v$,
\be
v=\sum_A \frac{ \bar{\rho}_A+\bar{p}_A}{\bar{\rho}+\bar{p}}\, v_A=\sum_A \frac{ 1+w_A}{1+w}\, v_A~.
\ee

As for the perturbed stress-energy tensor of the dark energy effective fluid, the only non-vanishing components are
\be
\delta \tilde{T}^0_{\pha 0}=-\delta \rho_x~,~~
\delta  \tilde{T}^i_{\;  j}=-\delta \rho_x \, \delta^i_{\;  j}~.\label{Eq:PerturbedDE}
\ee
The corresponding four-velocity is completely arbitrary%
\footnote{This statetement is only true for $w_x=-1$. For a comparison with other dark energy models with  $w_x\neq-1$, see Ref.~\cite{Valiviita:2008iv}.};
this can be seen by observing that all timelike vectors are eigenvectors of $\tilde{T}^a_{\pha b}$ with the same eigenvalue $-\delta \rho_x$. For this reason we will adopt the convention that the four-velocity of dark energy is given by the total four-velocity $u^a$.

Lastly, we recall the definitions of the gauge-invariant curvature perturbations on constant energy density hyper-surfaces \cite{Valiviita:2008iv, Malik:2008im}
\be\label{Eq:ZetaDef}
\zeta_A=-\psi-\mathcal{H}\frac{\delta \rho_A}{\bar{\rho}^{\prime}_A}~,\hspace{1em} \zeta=-\psi-\mathcal{H}\frac{\delta \rho}{\bar{\rho}^{\prime}}~,
\ee
defined for the fluid $A$ and the total energy density, respectively. The gauge-invariant entropy perturbation of the fluid $A$ relative to the fluid $B$ is defined as
\be\label{Eq:EntropyPertDef_General}
S_{AB}=3\mathcal{H}\left(\frac{\delta \rho_B}{\bar{\rho}^{\prime}_B}-\frac{\delta \rho_A}{\bar{\rho}^{\prime}_A}\right)=3(\zeta_A-\zeta_B)~,
\ee
which for barotropic fluids reduces to
\be\label{Eq:EntropyPertDef}
S_{AB}=\frac{\delta_A}{1+w_A}-\frac{\delta_B}{1+w_B}~.
\ee

\subsection{Perturbed field equations}

The perturbed field equations can be derived from Eq.~\eqref{Eq:EffectiveEFE}
\be\label{Eq:EffectiveEFE_Perturbed}
\delta G^{a}_{\pha b}=\kappa (\delta T^{a}_{\pha b}+\delta \tilde{T}^{a}_{\pha b})~.
\ee
From Eq.~\eqref{Eq:EffectiveEFE_Perturbed} written in components in the coordinate chart defined by Eq.~\eqref{Eq:perturbed_metric}, and using Eqs.~\eqref{Eq:PerturbedTmunu} and \eqref{Eq:PerturbedDE}, we obtain
\begin{subequations}
\begin{align}
\delta G^0_{\pha 0}&=2a^{-2}\left[-\triangle\psi +3\mathcal{H}\left(\mathcal{H}\phi+\psi^{\prime}\right)+\mathcal{H}\triangle(B-E^{\prime})     \right]=-\kappa(\delta \rho+\delta \rho_x) ~,\label{Eq:PertEinstein00}\\
\delta G^0_{\pha i}&=-2a^{-2}\pa_i\left(\mathcal{H}\phi+\psi^{\prime}\right)=\kappa(\bar{\rho}+\bar{p})\pa_i(v+B)~,\label{Eq:PertEinstein0i}\\
\delta G^i_{\pha 0}&=2a^{-2}\pa_i\left[\mathcal{H}\phi+\psi^{\prime}+\left(2\mathcal{H}^2-\frac{a^{\prime\prime}}{a}\right)B\right] =-\kappa(\bar{\rho}+\bar{p})\pa_i v~,\label{Eq:PertEinsteini0}
\end{align}
\begin{align}
\delta G^i_{\pha j}=a^{-2}\left\{ 2\left[\psi^{\prime\prime}+\mathcal{H}(\phi^{\prime}+2\psi^{\prime}) +\left(2\frac{a^{\prime\prime}}{a}-\mathcal{H}^2\right)\phi -\tfrac{1}{2}\triangle \mathcal{D}\right]\delta^i_{\; j}+\pa^i\pa_j \mathcal{D} \right\} \nonumber \\
=\kappa\left[(\delta p-\delta \rho_x) \delta^i_{\;  j} +\pi^i_{\; j}\right]~,\label{Eq:PertEinsteinij}
\end{align}
\end{subequations}
having defined the quantity
\be
\mathcal{D}=\psi-\phi+E^{\prime\prime}-B^{\prime}+2\mathcal{H}(E^{\prime}-B)~.
\ee
From Eqs.~\eqref{Eq:EffectiveEFE_Perturbed} we obtain the following set of independent equations, valid in any gauge
\begin{subequations}\label{Eq:PerturbedEFEall}
\begin{align}
- \triangle \psi +3 \mathcal{H}(\psi^{\prime}+ \mathcal{H}\phi) +\mathcal{H} \triangle (B-E^{\prime}) &=-\frac{ \kappa}{2}a^2(\delta \rho+\delta \rho_x)~,\label{Eq:PerturbedEFE1}\\
\mathcal{H}\phi+\psi^{\prime}&=-\frac{ \kappa}{2}a^2(\bar{\rho}+\bar{p})(v+B)~,\label{Eq:PerturbedEFE2}\\
\mathcal{D} &=\kappa\, a^{2} \pi ~,\label{Eq:PerturbedEFE3}
\end{align}
\be
\psi^{\prime\prime}+\mathcal{H}(\phi^{\prime}+2\psi^{\prime}) +\left(2\frac{a^{\prime\prime}}{a}-\mathcal{H}^2\right)\phi-\frac{1}{3}\triangle \mathcal{D} =\frac{\kappa}{2}a^2 \left(\delta p-\delta \rho_x \right)~.
\ee
\end{subequations}
It can be verified that the same set of equations can be derived directly from the original traceless field equations~\eqref{Eq:TracelessFieldEquations}.

\subsection{Longitudinal gauge}

The traceless field equations \eqref{Eq:TracelessFieldEquations} retain the full diffeomorphism invariance of general relativity. It is therefore convenient to exploit the gauge freedom and analyse the dynamics of perturbations in the longitudinal gauge $E=B=0$. In this gauge, the scalar perturbations $\phi$, $\psi$ are equal to the gauge-invariant Bardeen potentials $\Phi_{\rm B}$, $\Psi_{\rm B}$, respectively~\cite{Mukhanov:1990me}.
The perturbed field equations in this gauge are
\begin{subequations}\label{Eq:PerturbedEFEall_Longitudinal}
\begin{align}
- \triangle \psi +3 \mathcal{H}(\psi^{\prime}+ \mathcal{H}\phi) &=-\frac{ \kappa}{2}a^2(\delta \rho+\delta \rho_x)~,\\
\mathcal{H}\phi+\psi^{\prime}&=-\frac{ \kappa}{2}a^2(\bar{\rho}+\bar{p})v~,\\
\psi-\phi &=\kappa\, a^{2} \pi ~,
\end{align}
\be
\psi^{\prime\prime}+\mathcal{H}(\phi^{\prime}+2\psi^{\prime}) +\left(2\frac{a^{\prime\prime}}{a}-\mathcal{H}^2\right)\phi =\frac{\kappa}{2}a^2 \left(\delta p-\delta \rho_x +\frac{2}{3}\triangle\pi\right)~.
\ee
\end{subequations}

\subsection{Continuity equations for matter fields}

To first order in perturbation theory, projecting the continuity equation for the full stress-energy tensor of matter \eqref{Eq:ContinuityFullA} along the unperturbed total four-velocity $\bar{u}^a$ and its orthogonal subspace (using the projector $\bar{q}_a^{\; c}=\delta_a^{\; c}+\bar{u}_a \bar{u}^c$), respectively gives (in a generic gauge)
\begin{subequations}
\be 
\delta \rho^{\prime}_A+ 3\mathcal{H}  (\delta \rho_A+\delta p_A)+(\bar{\rho}_A+\bar{p}_A)\left(\triangle E^{\prime}+\triangle v_A-3 \psi^{\prime}\right)= \kappa^{-1}\delta Q^{\prime}_A~,\label{Eq:rho_matter_pert}
\ee
\be 
\left[(\bar{\rho}_A+\bar{p}_A) (B+v_A)\right]^{\prime} +(\bar{\rho}_A+\bar{p}_A) \Big[ \phi + 4 \mathcal{H} (B+v_A) \Big] + \delta p_A +\frac{2}{3}\triangle \pi_A =-\kappa^{-1}\delta Q_A~. \label{Eq:pressure_matter_pert} 
\ee
\end{subequations}
Analogous equations for the total quantities characterizing the fluid mixture can be obtained by summing Eqs.~\eqref{Eq:rho_matter_pert},~\eqref{Eq:pressure_matter_pert} over $A$ and using \eqref{Eq:PerturbedFluid},
\begin{subequations}
\begin{align}
\delta \rho^{\prime}+ 3\mathcal{H}  (\delta \rho+\delta p)+(\bar{\rho}+\bar{p})\left(\triangle E^{\prime}+\triangle v-3 \psi^{\prime}\right)&= \kappa^{-1} \delta Q^{\prime}=-\delta \rho_x^{\prime}~,\label{Eq:rho_matter_pert_TOT}  \\
\left[(\bar{\rho}+\bar{p}) (B+v)\right]^{\prime} +(\bar{\rho}+\bar{p}) \phi +\delta p +4  \mathcal{H} (\bar{\rho}+\bar{p}) (B+v)+\frac{2}{3}\triangle \pi&=-\kappa^{-1}\delta Q=\delta \rho_x~. \label{Eq:pressure_matter_pert_TOT} 
\end{align}
\end{subequations}
Note that the above equations are a particular case of the class of models studied in Ref.~\cite{Valiviita:2008iv}, a detailed comparison is given in Sec.~\ref{Sec:Comparison}.

It is also useful to introduce the velocity perturbation $\theta_A$ and density contrast $\delta_A$, defined as usual by
\be\label{Eq:DefThetaDelta}
\theta_A=\triangle (v_A+B)~,~~\delta_{A}\coloneqq \frac{\delta \rho_{A}}{\bar{\rho}_{A}}~.
\ee
The definitions of the corresponding quantities $\theta$ and $\rho$ for the total matter stress-energy tensor are entirely analogous; in particular, using Eq.~\eqref{Eq:PerturbedFluid}, we obtain
\be\label{Eq:ThetaTotal}
\sum_A (\bar{\rho}_A+\bar{p}_A) \theta_A=(\bar{\rho}+\bar{p}) \theta~.
\ee
For the dark energy fluid, similarly we define $\delta_x=\delta \rho_{x}/\bar{\rho}_{x}=\delta Q/\bar{Q}$~. 

The pressure perturbation can be decomposed into its adiabatic and non-adiabatic components,
\be\label{Eq:PressurePert}
\delta p_A=c_{a\,A}^2\delta\rho_A+ \delta p_{{\rm nad}\,A}=c_{a\,A}^2\delta\rho_A+ (c_{s\,A}^2-c_{a\,A}^2) \left[\delta \rho_A +\bar{\rho}_A^{\prime}(v_A+B)\right]~,
\ee
where the adiabatic speed of sound is defined as
\be\label{Eq:AdiabaticSoundSpeed}
c_{a\,A}^2\coloneqq\frac{\bar{p}_A^{\prime}}{\bar{\rho}_A^{\prime}}=w_{A}+\frac{w_A^{\prime}}{\bar{\rho}_A^{\prime}/\bar{\rho}_A}~,
\ee
while $c_{s\,A}$ is the speed of sound in the fluid rest frame
\be\label{Eq:RestFrameSoundSpeed}
c_{s\,A}^2\coloneqq \left.\frac{\delta p_A}{\delta \rho_A}\right|_{\rm rest\; frame}~.
\ee
Using the definition~\eqref{Eq:DefThetaDelta} for $\theta_A$, we can rearrange Eq.~\eqref{Eq:PressurePert} and write it in Fourier space as
\be\label{Eq:PressureFluctuation}
\delta p_A=c_{s\,A}^2\delta\rho_A-(c_{s\,A}^2-c_{a\,A}^2)\bar{\rho}_A^{\prime}\frac{\theta_A}{k^2} ~.
\ee
We remark that equation~\eqref{Eq:PressureFluctuation} includes the effects of energy-momentum transfer through $\bar{\rho}_A^{\prime}$ and is equivalent to Eq.~(29) in Ref.~\cite{Valiviita:2008iv}.

The evolution equations for $\delta_{A}$ and $\theta_A$ in Fourier space are
\begin{subequations}\label{Eq:MatterPerturbationsDeltaTheta}
\begin{align} \label{Eq:DeltaContinuity}
\delta^{\prime}_A+& \left(3\mathcal{H}  (c_{s\,A}^2-w_A ) + \kappa^{-1}  \frac{\bar{Q}^{\prime}_A}{\bar{\rho}_A}\right)\delta_A+(w_A+1)\theta_A-3(w_A+1)\psi^{\prime} +(w_A+1)k^2(B-E^{\prime}) \nonumber \\
& \qquad +3\mathcal{H}  (c_{s\,A}^2-c_{a\,A}^2) \left[3\mathcal{H}  (1+w_A) - \kappa^{-1} \frac{\bar{Q}_A^{\prime}}{\bar{\rho}_A}\right] \frac{\theta_A}{k^2} = \kappa^{-1}  \frac{\delta Q^{\prime}_A}{\bar{\rho}_A}~,
\end{align}
\begin{align} \label{Eq:ThetaContinuity}
\theta^{\prime}_A+{\cal H} (1-3c_{s\,A}^2)\theta_A\,-k^2\phi\,&-\frac{k^2}{1+w_A}c_{s\,A}^2\delta_A+\frac{2k^4}{3(1+w_A)\bar{\rho}_A}\pi_A \nonumber \\ &=\kappa^{-1}\left[\frac{k^2}{\bar{\rho}_A(1+w_A)}\delta Q_A -\left(\frac{1+c_{s\, A}^2}{1+w_A}\right)\frac{\bar{Q}^{\prime}_A}{\bar{\rho}_A}\theta_A \right] ~.
\end{align}
\end{subequations}

We stress that there is no additional dynamical equation for $\delta_x$, which is just given by $\delta_x=\delta Q/\bar{Q}$~. To determine the evolution of $\delta_x$, it is necessary to assume a model for the energy-momentum transfer potential, as discussed in Section~\ref{Sec:TransferModel}. (As an aside, formally the sound speed of dark energy could also be defined using Eqs.~\eqref{Eq:AdiabaticSoundSpeed} and \eqref{Eq:RestFrameSoundSpeed} giving $c_s^2=c_a^2=w_x=-1$, but these quantities do not appear in any dynamical equations. Similarly, one could apply Eq.~\eqref{Eq:DeltaContinuity} to dark energy by substituting $c_s^2=c_a^2=w_x=-1$ and $Q_x=-Q$, which gives the identity $\delta^{\prime}_x=(\delta Q/\bar{Q})^{\prime}$. Equation~\eqref{Eq:ThetaContinuity} cannot be applied for $w_x=-1$, but Eq.~\eqref{Eq:pressure_matter_pert} with $\pi=0$ gives the identity $\delta p_x=-\kappa^{-1}\delta Q_x =-\delta \rho_x$.)

Finally, note that these equations may need to be further modified in the presence of other interactions independent from the $Q_A$; for instance, baryons and photons interact via Thomson scattering before recombination, which leads to an extra term in Eq.~\eqref{Eq:ThetaContinuity} for those matter components~\cite{Ma:1995ey}.

\subsection{Comparison with other interacting dark-sector models}
\label{Sec:Comparison}

Let us now compare our model with Ref.~\cite{Valiviita:2008iv}, where a general energy-momentum transfer is decomposed with respect to the total four-velocity as
\be
\kappa^{-1} J_a^A=\mathcal{Q}^A u_a+\mathcal{F}^A_a~,~~\mbox{with}~~ u^a\mathcal{F}^A_a=0~.
\ee
and the momentum density transfer rate is assumed to arise from a potential: $\mathcal{F}^A_a=a(0,\pa_i f_A)$. Given Eq.~\eqref{Eq:DefJ_A}, our model can be obtained as a particular case for
\be\label{Eq:Correspondence}
\mathcal{Q}^A=\kappa^{-1}\, u^a \nabla_a Q^A~,~~\mathcal{F}^A_a=-\kappa^{-1}\, q_a^{\; c}\nabla_c Q^A~,
\ee
where we introduced the projector $q_a^{\; c}=\delta_a^{\; c}+u_a u^c$.  More explicitly, expanding Eq.~\eqref{Eq:Correspondence} we obtain the following relations: at the background level
\be\label{Eq:Correspondence2}
\bar{\mathcal{Q}}_A=\kappa^{-1}\,a^{-1}\bar{Q}_A^{\prime}~,
\ee
and for first-order scalar perturbations
\be\label{Eq:Correspondence3}
\delta \mathcal{Q}_A=\kappa^{-1}\,a^{-1}\left(\delta Q_A^{\prime}-\phi\,  \bar{Q}_A^{\prime}\right)~,~~\mathcal{F}^A_i=-\kappa^{-1}\,\pa_i\left[\delta Q_A+(v+B) \bar{Q}_A^{\prime}\right]~.
\ee
Combining the first equation in \eqref{Eq:Correspondence3} with \eqref{Eq:Correspondence2},
\be\label{Eq:Correspondence3bis}
a\left( \delta \mathcal{Q}_A+\phi  \bar{\mathcal{Q}}_A \right)=\kappa^{-1}\,\delta Q_A^{\prime} ~.
\ee
The latter equation in~\eqref{Eq:Correspondence3}, combined with \eqref{Eq:Correspondence2}, implies that the potential $f_A$ satisfies
\be\label{Eq:Correspondence4}
a\left[ f_A+(v+B)\bar{\mathcal{Q}}_A\right]=-\kappa^{-1}\, \delta Q_A ~.
\ee
Equations~\eqref{Eq:Correspondence3bis} and \eqref{Eq:Correspondence4} show that \eqref{Eq:rho_matter_pert} and \eqref{Eq:pressure_matter_pert} are indeed a particular case of Eqs.~(20) and (21) in Ref.~\cite{Valiviita:2008iv}.

\section{Model for the energy-momentum transfer potential}
\label{Sec:TransferModel}

In order to close the system of equations for gravitational and matter perturbations, Eqs.~\eqref{Eq:PerturbedEFEall_Longitudinal} and \eqref{Eq:MatterPerturbationsDeltaTheta}, we need to assume a model for the energy-momentum transfer. In principle, this should come from fundamental physics, as suggested for instance in Refs.~\cite{Josset:2016vrq,Perez:2017krv}. However, here we adopt a simple phenomenological model. Specifically, we assume that the non-conservation of the total stress-energy of matter is entirely due to cold dark matter and consider the following model for the covariant energy-momentum transfer
\be\label{Eq:DiffusionModelJ}
J_a=- \epsilon\, \kappa\, \nabla_a T_{uu}^{\rm\scr CDM}~,
\ee
where $T_{uu}^{\rm\scr CDM}\coloneqq u^a u^b T_{ab}^{\rm\scr CDM}$ is the energy density of cold dark matter in its local rest frame, and $\epsilon$ is a dimensionless coupling constant, assumed to be small $|\epsilon|\ll1$. From Eq.~\eqref{Eq:DiffusionModelJ} and the definition \eqref{Eq:DefQ},
\be\label{Eq:DiffusionModelQ}
Q=-\Lambda_{\rm f}+\epsilon\, \kappa\, T_{uu}^{\rm\scr CDM}=-\Lambda_{\rm f}+\epsilon\, \kappa\,\rho_c~,
\ee
where $\Lambda_{\rm f}$ is an integration constant, giving the value of the cosmological constant in the distant future.
This is a covariant form of a model previously proposed in Ref.~\cite{Perez:2020cwa}. At the background level we have
\be\label{Eq:DiffusionBckgnd}
\bar{Q}=-\Lambda_{\rm f}+\kappa\, \epsilon\, \bar{\rho}_c ~,
\ee
which implies, using Eq.~\eqref{Eq:ContinuityBckgrnd}
\be\label{Eq:InteractingDM_BackgroundEnergyDensity}
\bar{\rho}^{\prime}_c+3\mathcal{H}\bar{\rho}_c=\epsilon\, \bar{\rho}_c^{\prime} ~\implies~ \bar{\rho}_c(\eta)=\bar{\rho}_{c,i} \left(\frac{a_{i}}{a(\eta)}\right)^{\frac{3}{1-\epsilon}}~,
\ee
where $\bar{\rho}_{c,i}$ and $a_i$ are initial conditions at $\eta_i$~. Equation~\eqref{Eq:InteractingDM_BackgroundEnergyDensity} shows that interacting dark matter has an effective barotropic equation of state
\be
w_{\rm eff}=\frac{\epsilon}{1-\epsilon}~.
\ee
Expanding Eq.~\eqref{Eq:DiffusionModelQ} to first order in perturbation theory, we obtain
\be\label{Eq:DiffusionPert}
\delta Q=\kappa\, \epsilon\, \delta \rho_c= \kappa\, \epsilon\, \bar{\rho}_c\, \delta_c=\kappa\, \epsilon\, \bar{\rho}_{c,i} \left(\frac{a_{i}}{a(\eta)}\right)^{\frac{3}{1-\epsilon}} ~ \delta_c ~.
\ee

In the radiation-dominated era, $\mathcal{H}=1/\eta$ and $a(\eta)= a_i\, \eta/\eta_i$, so Eqs.~\eqref{Eq:DiffusionBckgnd}, \eqref{Eq:DiffusionPert} become
\be\label{Eq:QinRDE}
\bar{Q}=-\Lambda_{\rm f}+\kappa\, \epsilon\, \bar{\rho}_{c,i} \left(\frac{\eta_{i}}{\eta}\right)^{\frac{3}{1-\epsilon}} ~,\hspace{1em}
\delta Q=\kappa\, \epsilon\, \bar{\rho}_{c,i} \left(\frac{\eta_{i}}{\eta}\right)^{\frac{3}{1-\epsilon}} \delta_c~.
\ee
From $\bar{Q}$ and $\delta Q$ we can compute the energy density and density contrast of dark energy using Eq.~\eqref{Eq:DefRhoX}
\be \label{Eq:DEsol_sub1}
\bar{\rho}_x= \kappa^{-1}\Lambda_{\rm f}- \epsilon\, \bar{\rho}_c=\kappa^{-1}\Lambda_{\rm f}-\epsilon\, \bar{\rho}_{c,i} \left(\frac{a_{i}}{a(\eta)}\right)^{\frac{3}{1-\epsilon}}~,
\ee
\be \label{Eq:DEsol_sub2}
\delta \rho_x= -\epsilon\, \bar{\rho}_c\, \delta_c=-\epsilon\, \bar{\rho}_{c,i} \left(\frac{a_{i}}{a(\eta)}\right)^{\frac{3}{1-\epsilon}} \delta_c~.  
\ee
Both of these two equations hold generally; it is only the dynamics of $a(\eta)$ which depend on the era---for example, during radiation-domination $a(\eta) \propto \eta$. From Eq.~\eqref{Eq:DEsol_sub2} it is clear that dark energy perturbations are well-behaved as long as the energy density of dark matter evolves regularly, as shall be shown in the next section.

As a consequence of Eqs.~\eqref{Eq:DEsol_sub2}, the entropy perturbation of dark energy relative to dark matter is always zero
\be
\zeta_x=-\psi -\mathcal{H}\frac{\delta \rho_x}{\bar{\rho}_x^{\prime}}=-\psi -\mathcal{H}\frac{\delta \rho_c}{\bar{\rho}_c^{\prime}}=\zeta_c ~\implies~ S_{xc}=3(\zeta_x-\zeta_c)=0~.
\ee
This is not surprising, since the energy-momentum transfer potential in Eq.~\eqref{Eq:DiffusionModelJ} is such that it follows the evolution of cold dark matter adiabatically.

\section{Radiation domination}
\label{Sec:RDE}

We study the evolution of metric and matter perturbations during the radiation-dominated era before recombination ($\mathcal{H}=\eta^{-1}$). Throughout this section we assume tight coupling, whereby baryons and photons have the same velocity perturbation $\theta_b=\theta_\gamma$ and therefore can be described as components of a single entity, the baryon-photon fluid.

We start by particularizing Eqs.~\eqref{Eq:DeltaContinuity}, \eqref{Eq:ThetaContinuity} to describe the different matter species. For cold dark matter we have
\begin{subequations}\label{Eq:ContinuityEquations_RDE}
\be
\delta^{\prime}_c+\theta_c-3\psi^{\prime}+k^2(B-E^{\prime})
= \kappa^{-1}\left(\frac{\delta Q^{\prime}}{\bar{\rho}_c}-\frac{\bar{Q}^{\prime}}{\bar{\rho}_c}\delta_c\right)=\epsilon\, \delta_c^{\prime}~,
\ee
\be
\theta^{\prime}_c+{\cal H} \theta_c-k^2\phi=\kappa^{-1}\left(\frac{k^2}{\bar{\rho}_c}\delta Q -\frac{\bar{Q}^{\prime}}{\bar{\rho}_c}\theta_c \right)=\epsilon\, k^2 \delta_c +\frac{3\epsilon}{1-\epsilon}\mathcal{H} \theta_c~,
\ee
\end{subequations}
where in the second equalities we used Eqs.~\eqref{Eq:DiffusionBckgnd} and \eqref{Eq:DiffusionPert}.

Photons obey the relativistic equation of state $w=\frac{1}{3}$ and have zero non-adiabatic pressure, which implies $c_{s}^2=c_{a}^2=\frac{1}{3}$. Their anisotropic stress is negligible due to Thomson scattering~\cite{Ma:1995ey}. For baryons we assume $w_b\ll 1$ and vanishing non-adiabatic pressure perturbation, so that the sound-speed is non-relativistic and given by $c_{s,b}^2=w_b$\,; also in this case there is no anisotropic stress%
\footnote{While the adiabatic sound speed of baryons can be consistently neglected in the energy balance equation, its contribution needs to be included in the momentum balance equation \eqref{Eq:RDEcontinuity_baryons} because it affects the evolution of  large-$k$ sub-horizon modes \cite{Ma:1995ey}.}.
For photons and baryons the energy-momentum balance equations~\eqref{Eq:DeltaContinuity}, \eqref{Eq:ThetaContinuity}, supplemented by the Thomson scattering interaction term~\cite{Ma:1995ey}, are
\be\label{Eq:RDEcontinuity_photons}
\delta^{\prime}_\gamma+\frac{4}{3}\theta_{\gamma}-4\psi^{\prime}+\frac{4}{3}k^2(B-E^{\prime}) =0~,\hspace{1em}
\theta_\gamma^{\prime}-k^2\phi-\frac{k^2}{4}\delta_\gamma= \tau_c^{-1}  (\theta_b - \theta_\gamma)~,
\ee
\be\label{Eq:RDEcontinuity_baryons}
\delta^{\prime}_b+\theta_b-3\psi^{\prime} +k^2(B-E^{\prime})= 0~,\hspace{1em}
\theta_b^{\prime}+{\cal H}\theta_b-k^2 c_{s,b}^2\delta_b-k^2\phi=R\,\tau_c^{-1} (\theta_\gamma-\theta_b)~,
\ee
where $R=\frac{4\bar{\rho}_\gamma}{3\bar{\rho}_b}$ and $\tau_c=(a n_e \sigma_{ T})^{-1}$. From Ref.~\cite{Ma:1995ey} we have the slip equation
\be
\theta_b^{\prime}-\theta_\gamma^{\prime}=\frac{2R}{1+R}\mathcal{H}(\theta_b-\theta_\gamma)+\mathcal{O}(\tau_c)~,
\ee
ensuring the consistency of the tight-coupling condition $\theta_\gamma=\theta_b$ at early times during the radiation-dominated era. For this reason, we drop the $\theta_b^{\prime}$ equation from this point forward and only keep the $\theta_\gamma^{\prime}$ equation, setting $\theta_\gamma=\theta_b$ (giving zero on the right of the equation.).

Light neutrinos (approximated as massless, with $w_\nu=\tfrac{1}{3}$) are governed by equations similar to \eqref{Eq:RDEcontinuity_photons} but allow for a non-zero anisotropic stress~\cite{Ma:1995ey}
\be\label{Eq:RDEcontinuity_neutrinos}
\delta^{\prime}_\nu+\frac{4}{3}\theta_{\nu}-4\psi^{\prime} +\frac{4}{3}k^2(B-E^{\prime})= 0~,\hspace{1em}
\theta_\nu^{\prime}-k^2\phi-\frac{k^2}{4}\delta_\nu+k^2\sigma_\nu =0 ~,
\ee
having defined $\sigma_{\nu}=\left(2 k^2/3(\bar{\rho}_\nu+\bar{p}_\nu)\right)\pi_\nu=\left(k^2/2\bar{\rho}_\nu\right)\pi_\nu$~.
In order to close the system, the continuity equations~\eqref{Eq:RDEcontinuity_neutrinos} need to be complemented by the evolution equation for $\sigma_{\nu}$~, obtained from the quadrupolar moment of the Boltzmann equation~\cite{Ma:1995ey} (we neglect the neutrino octopole term, following Ref.~\cite{Valiviita:2008iv})
\be
\sigma_\nu^{\prime}=\frac{4}{15}\theta_\nu~.
\ee

\subsection{Super-horizon evolution}

Next, we focus on super-horizon modes with $k\ll \mathcal{H}$ in the tight-coupling approximation $\theta_b=\theta_\gamma$. We fix the longitudinal gauge $E=B=0$ for convenience. The system of equations we need to solve in this regime comprises the energy-momentum balance equations for matter 
\begin{subequations}\label{Eq:ContinuitySuperHubble}
\begin{align}
 &\delta^{\prime}_\gamma+\frac{4}{3}\theta_{\gamma}-4\psi^{\prime} =0~,\hspace{1em}
\theta_\gamma^{\prime}-k^2\phi-\frac{k^2}{4}\delta_\gamma= 0~,\hspace{1em}
\delta^{\prime}_b+\theta_\gamma-3\psi^{\prime}= 0~,\\
&(1-\epsilon)\delta^{\prime}_c+\theta_c-3\psi^{\prime}
= 0~,\hspace{1em}
\theta^{\prime}_c+\left(\frac{1-4\epsilon}{1-\epsilon}\right)\eta^{-1}\theta_c-k^2\phi-\epsilon\, k^2\delta_c=0~,\\
&\delta^{\prime}_\nu+\frac{4}{3}\theta_{\nu}-4\psi^{\prime} = 0~,\hspace{1em}
\theta_\nu^{\prime}-k^2\phi-\frac{k^2}{4}\delta_\nu+k^2\sigma_\nu
=0 ~,\hspace{1em}
\sigma_\nu^{\prime}=\frac{4}{15}\theta_\nu~,
\end{align}
\end{subequations}
along with the gravitational field equations~\eqref{Eq:PerturbedEFEall_Longitudinal}
\begin{subequations}\label{Eq:PerturbedEFEallRDE}
\begin{align}
k^2 \psi +3 \eta^{-1}(\psi^{\prime}+ \eta^{-1}\phi) &=-\frac{ \kappa}{2}\left(\frac{a_i}{\eta_i}\right)^2\eta^2(\delta \rho+\delta \rho_x)~,\label{Eq:PerturbedEFE1RDE}\\
\eta^{-1}\phi+\psi^{\prime}&=\frac{ \kappa}{2}\left(\frac{a_i}{\eta_i}\right)^2 \eta^2 k^{-2}(\bar{\rho}+\bar{p})\theta~,\label{Eq:PerturbedEFE2RDE}\\
\psi-\phi&=\kappa\,\left(\frac{a_i}{\eta_i}\right)^2 \eta^{2}\, \left(\frac{2\bar{\rho}_\nu}{k^2}\right)\sigma_\nu ~,\label{Eq:PerturbedEFE3RDE}
\end{align}
\be\label{Eq:PerturbedEFE4RDE}
\psi^{\prime\prime}+\eta^{-1}(\phi^{\prime}+2\psi^{\prime}) -\eta^{-2}\phi =\frac{\kappa}{2}\left(\frac{a_i}{\eta_i}\right)^2 \eta^2 \left(\delta p-\delta \rho_x -\frac{4}{3}\bar{\rho}_\nu\, \sigma_\nu  \right)~.
\ee
\end{subequations}

As in Ref.~\cite{Valiviita:2008iv}, we look for the leading-order behaviour of super-horizon perturbations ($k\eta\ll 1$) with the following power-law ans{\"a}tze
\begin{subequations}
\begin{align}
&\psi=A_{\psi}(k\eta)^{n_\psi}~,\hspace{1em} \phi=A_{\phi}(k\eta)^{n_\phi}~,\\
&\delta_A=B_A (k\eta)^{n_A}~,\hspace{1em} \theta_A=C_A (k\eta)^{s_A}~,\hspace{1em} \sigma_\nu=D_\nu (k\eta)^{n_\sigma}~.
\end{align}
\end{subequations}
All solutions will be determined up to an arbitrary rescaling of $A_{\psi}$.
 
We start by solving the $\delta_A^{\prime}$ equations in \eqref{Eq:ContinuitySuperHubble}, which give to leading order
\be\label{Eq:DensityConstrastSol}
\delta_\gamma=4 \psi+K_\gamma~,~~\delta_b=3 \psi+K_b~,~~\delta_\nu=4 \psi+K_\nu~,~~\delta_c=\frac{3}{1-\epsilon} \psi+K_c~,
\ee
where the $K_A$ are integration constants. Equation~\eqref{Eq:PerturbedEFE3RDE} implies $n_\psi=n_\phi$, that is
\be
\phi=J\, \psi~,\hspace{1em}\mbox{with}~~J\coloneqq \frac{A_\phi}{A_\psi}~.
\ee

\subsection{Constant mode (for $n_\psi= 0$)}

The super-horizon solution corresponding to $n_\psi = 0$ is the constant mode; in this case Eq.~\eqref{Eq:PerturbedEFE1RDE} (whose right side is dominated by radiation) gives
\be
\delta_{\rm rad}=-2\phi ~,
\ee
which in turn implies
\be
(1-\Omega_\nu)\, \delta_\gamma+  \Omega_\nu\, \delta_\nu=-2\phi ~.
\ee
Thus, the two integration constants in Eq.~\eqref{Eq:DensityConstrastSol} must satisfy
\be\label{Eq:IntConst_NuGamma}
(1-\Omega_\nu)\, K_\gamma  +  \Omega_\nu\, K_\nu=-2(J+2)\psi
\ee
Recalling the definition of the gauge-invariant entropy perturbation~\eqref{Eq:EntropyPertDef} of neutrinos relative to photons
\be\label{Eq:Entropy_NuGamma}
S_{\nu \gamma}=\frac{3}{4}(\delta_\nu-\delta_\gamma)=\frac{3}{4}(K_\nu-K_\gamma) ~,
\ee
we solve Eqs.~\eqref{Eq:IntConst_NuGamma}, \eqref{Eq:Entropy_NuGamma} to express the integration constants $K_\gamma$, $K_\nu$ in terms of $\Omega_\nu$ and $S_{\gamma b}$~. Substituting this back in Eq.~\eqref{Eq:DensityConstrastSol} gives
\be\label{Eq:DensityContrast_RadSol}
\delta_\gamma=-2\phi-\frac{4}{3}S_{\nu\gamma}\Omega_\nu~,~~  \delta_\nu=-2\phi+\frac{4}{3}S_{\nu\gamma}(1-\Omega_\nu)~.
\ee
The density contrast of baryons and cold dark matter can then be obtained using the photon solution~\eqref{Eq:DensityContrast_RadSol} and recalling the definition of their gauge-invariant entropy perturbations relative to baryons \eqref{Eq:EntropyPertDef_General}
\be
S_{\gamma b}=\frac{3}{4}\delta_\gamma-\delta_b
~,\hspace{1em} S_{c b}=(1-\epsilon)\delta_c-\delta_b~.
\ee
Thus, the solutions for the density contrast for baryons and cold dark matter read
\be\label{Eq:DensityConstrastSol_constant}
\delta_b=-\frac{3}{2} \phi-S_{\gamma b}-S_{\nu\gamma}\Omega_\nu~,\hspace{1em}  \delta_c=-\frac{3}{2(1-\epsilon)} \phi+\frac{1}{1-\epsilon}(S_{cb}-S_{\gamma b}-S_{\nu\gamma}\Omega_\nu)~.
\ee
 
With this, the velocity perturbations and the anisotropic stress are given by
\begin{subequations}
\begin{align}
&\theta_\gamma=\left(\frac{1}{2}\phi-\frac{1}{3}S_{\nu\gamma}\Omega_\nu\right)k(k\eta)~,\hspace{1em} \theta_\nu=\left(\frac{1}{2}\phi+\frac{1}{3}S_{\nu\gamma}(1-\Omega_\nu)\right)k(k\eta)~,\\
&\theta_c=\left[\frac{1}{2}\phi+\frac{\epsilon}{2-5\epsilon}(S_{cb}-S_{\gamma b}-S_{\nu\gamma}\Omega_\nu)\right]k(k\eta)~,\\
&\sigma_\nu=\frac{1}{15}\left(\phi+\frac{2}{3}S_{\nu\gamma}(1-\Omega_\nu)\right)(k\eta)^2~.
\end{align}
\end{subequations}
The gravitational potentials $\psi$ and $\phi$ are both constant; the relation between them is obtained by solving Eq.~\eqref{Eq:PerturbedEFE3RDE}
\be
\psi=\left(1+\frac{2}{5}\Omega_\nu\right)\phi+\frac{4}{15}S_{\nu\gamma}\Omega_\nu (1-\Omega_\nu)~.
\ee
Thus, this super-horizon mode depends on four constants $\phi$, $S_{\nu\gamma}$, $S_{\gamma b}$, $S_{cb}$ (determined by the initial conditions) and on the neutrino background density parameter $\Omega_\nu$. Adiabatic initial conditions correspond to the choice $S_{\nu\gamma}=S_{\gamma b}=S_{cb}=0$.

The constant super-horizon solution obtained in this section can be decomposed, as usual, into a superposition of one adiabatic curvature perturbation and three independent isocurvature density perturbations (for photons, neutrinos and cold dark matter, relative to baryons), following, e.g., Refs.~\cite{Bucher:1999re, Langlois:2003fq}.

\subsection{Decreasing mode (for $n_\psi\neq 0$)}

For solutions with $n_\psi \neq 0$, since we are interested in growing or decaying modes we set the integration constants in Eq.~\eqref{Eq:DensityConstrastSol} all to zero,
\be\label{Eq:IntConst_DecayingMode}
K_\gamma=K_b=K_\nu=K_c=0~.
\ee
With this, it is straightforward to find the leading-order solutions to the momentum-balance equations in Eq.~\eqref{Eq:ContinuitySuperHubble}, obtaining
\be\label{Eq:SolThetas}
\theta_\gamma=\theta_\nu=\frac{J+1}{n_{\psi}+1}(k\eta)k\psi~,\hspace{1em} \theta_c=\frac{J+\frac{3\epsilon}{1-\epsilon}}{n_{\psi}+\frac{2-5\epsilon}{1-\epsilon}}(k\eta)k\psi=\frac{n_{\psi}+1}{n_{\psi}+\frac{2-5\epsilon}{1-\epsilon}}\frac{J+\frac{3\epsilon}{1-\epsilon}}{J+1}\theta_\gamma~,
\ee
and the anisotropic stress is
\be
\sigma_\nu=\frac{4}{15(n_\psi+2)}(k\eta)\frac{\theta_\nu}{k}~.
\ee
Next, we can solve Eq.~\eqref{Eq:PerturbedEFE3RDE} to determine $J$
\be\label{Eq:Jsol}
J=1-\frac{16\,\Omega_{\nu}}{5(n_\psi+1)(n_\psi+2)+8\,\Omega_{\nu}}~,
\ee
where $\Omega_{\nu}$ represents the neutrino energy density parameter (which is constant during the radiation dominated era)
\be
\Omega_{\nu}=\frac{\kappa\, a^2 \bar{\rho}_{\nu}}{3\mathcal{H}^2}=\frac{\bar{\rho}_{\nu}}{\bar{\rho}_{\gamma}+\bar{\rho}_{\nu}}~.
\ee
Lastly, we solve Eq.~\eqref{Eq:PerturbedEFE1RDE} in the super-horizon limit (and only considering the dominant contribution to $\delta\rho$ from radiations, i.e., photons and neutrinos), which gives
\be\label{Eq:nPsiSol}
n_\psi=-(2+J)~.
\ee

Combining equations \eqref{Eq:nPsiSol}, \eqref{Eq:Jsol} gives as possible solutions for $n_\psi$ and $J$
\be 
(n_\psi,J) = (-1,-1)~; \qquad (n_\psi,J) = \Bigg(\f{-5 \pm \sqrt{1 - \frac{32}{5} \Omega_\nu}}{2} ,   \f{1 \mp \sqrt{1 - \frac{32}{5} \Omega_\nu}}{2} \Bigg)~,
\ee
which clearly shows that $n_\psi$ remains bounded, contrary to what occurs for IDE models considered in \cite{Valiviita:2008iv}. Of these, the only solution that is continuous in the limit of zero anisotropic stress (i.e., $\Omega_\nu\to0$) is
\be
(n_\psi,J) =\Bigg(\f{-5 - \sqrt{1 - \tfrac{32}{5} \Omega_\nu}}{2} ,   \f{1 + \sqrt{1 - \tfrac{32}{5} \Omega_\nu}}{2} \Bigg)~.
\ee
In the limit that anisotropic stress is negligible and $\psi=\phi$, then $n_\psi = -3$, while for small $\Omega_\nu$ we have $n_\psi \approx -3 + \tfrac{8}{5} \Omega_\nu$ and $J\approx 1-\tfrac{8}{5}\Omega_\nu$~.  In all cases, including for large $\Omega_\nu$\,, $\textrm{Re}(n_\psi) < 0$ so large-scale non-adiabatic modes in $\psi$ do not blow up, rather they decrease in amplitude during the radiation-dominated era.

As for the dark energy fluid, we can obtain an explicit solution for $\delta_x$ by combining Eq.~\eqref{Eq:DEsol_sub2} with the solution \eqref{Eq:DensityConstrastSol} for $\delta_c$,
\be
\delta_x=-3\frac{\epsilon}{1-\epsilon} \left(\frac{\bar{\rho}_c}{\bar{\rho}_x} \right)\psi~.
\ee
Since both $\psi$ and $\bar{\rho}_c/\bar{\rho}_x$ are decreasing in time during the radiation dominated era%
\footnote{We have for the time derivative $\left(\bar{\rho}_c/\bar{\rho}_x\right)^{\prime}=\kappa^{-1}\Lambda_{\rm f}\bar{\rho}_c^{\prime}/\bar{\rho}_x^2<0$ at all times, assuming $\Lambda_{\rm f}>0$ to match late-time observations.},
$\delta_x$ is also decreasing. For given initial conditions on $\bar{\rho}_c$, the upper bound on $\delta_x$ can be made arbitrarily small by an appropriate choice of the value of $\epsilon$. As discussed in Section~\ref{Sec:Pert-T}, the velocity perturbation of dark energy is identified with that of the total fluid, $\theta_x=\theta$. We have, recalling Eq.~\eqref{Eq:ThetaTotal} and using the solution \eqref{Eq:SolThetas} for photons and neutrinos (sub-dominant matter species can be neglected), that $\theta\simeq \theta_\gamma$. Therefore, $\theta$ is also a decreasing function of time and the dark energy perturbations are regular.

Using the solution \eqref{Eq:DensityConstrastSol} with the integration constants \eqref{Eq:IntConst_DecayingMode}, we find that the gauge-invariant curvature perturbations~\eqref{Eq:ZetaDef} for the different matter components are all vanishing, $\zeta_A=0$~.
Therefore, also the relative entropy perturbations~\eqref{Eq:EntropyPertDef} trivially vanish, $S_{AB}=0$ for all pairs of fluids. This mode can be fully characterized by the gauge-invariant relative velocity perturbation $\theta_c-\theta_\gamma$, whose magnitude decreases with time.

\subsection{Section summary}
\label{Sec:Dsummary}

We conclude this section with a brief summary. During the radiation dominated era and in the tight-coupling regime, super-horizon perturbations are found to be either decaying or constant. Thus, the qualitative behaviour of scalar perturbations is quite similar to standard cosmology, with small departures in the behaviour of cold dark matter perturbations and with an evolving dark energy component $\rho_x$ that evolves adiabatically, both controlled by a dimensionless parameter $\epsilon$. There is no instability for linear perturbations in this regime for arbitrary choices of initial conditions, which marks a clear difference between our model and many other interacting dark-sector models, including the ones examined in Ref.~\cite{Valiviita:2008iv}. In the model of Ref.~\cite{Valiviita:2008iv}, the term driving the instability is the contribution of the energy-momentum transfer to the non-adiabatic pressure perturbation of the dark energy fluid; this is absent in our model, where we have exactly $\delta p_x=-\delta \rho_x$~. Importantly, since here $\theta_x = \theta$ there is no equation for the velocity perturbation of dark energy $\theta_x$ (and none is needed), which in Ref.~\cite{Valiviita:2008iv} grows unbounded.  Also, for the super-horizon modes considered here the $\epsilon\to0$ limit is continuous, and this limit gives the standard ${\rm \Lambda CDM}$ cosmology.

Finally, the instability is also absent if dark matter has a small non-adiabatic pressure perturbation. Although this gives extra terms in the continuity equations \eqref{Eq:ContinuityEquations_RDE}, these terms do not generate any instability either.

\section{Interacting dark matter domination}
\label{Sec:GMDE}

In this section we show that, due to interactions with dark energy, cold dark matter effectively behaves as a perfect fluid with a non-zero equation of state depending on the small parameter $\epsilon$. In fact, there is a general connection with the phenomenological generalized dark matter model (GDM) \cite{Hu:1998kj}, which has also been studied in Refs.~\cite{Thomas:2016iav, Kopp:2016mhm, Ilic:2020onu}. GDM is parametrised by three physical quantities, namely the equation of state, sound speed, and viscosity, which are in general time dependent. For the specific transfer potential considered in this paper, we obtain GDM with a constant equation of state, vanishing intrinsic entropy perturbation, and zero viscosity. Although this model has been previously considered in Ref.~\cite{Kopp:2016mhm}, here we focus specifically on the $\epsilon\to0$ limit in which $\Lambda$CDM is recovered, showing that this is a singular limit for sub-horizon modes.

Although we study the era of dark matter domination in this section, the GDM interpretation (and also the values of the effective equation of state and of the sound speed) is valid at all times.

The background evolution is obtained by solving the modified Friedman equation \cite{Perez:2020cwa}
\be\label{Eq:FriedmannMDE1}
\mathcal{H}^2=\frac{\kappa}{3}a^2(\bar{\rho}_c+\bar{\rho}_x)=\frac{1}{3}a^2\Big(\Lambda_{\rm f}+(1-\epsilon)\kappa\, \bar{\rho}_{c}\Big) ~,
\ee
with
\be\label{Eq:DMdensity}
\bar{\rho}_c(\eta)=\bar{\rho}_{c,i} \left(\frac{a_{i}}{a(\eta)}\right)^{\frac{3}{1-\epsilon}}~.
\ee
Equation~\eqref{Eq:DMdensity} already shows that interacting cold dark matter has an effective equation of state $w_{c,\rm eff}=\epsilon/(1-\epsilon)$~. The gravitational coupling is also effectively rescaled to $\kappa_{\rm eff}=(1-\epsilon)\kappa$.
During dark matter domination, $\Lambda_{\rm f}$ can be disregarded in the right side of Eq.~\eqref{Eq:FriedmannMDE1},
\be\label{Eq:FriedmannMDE2}
\mathcal{H}^2\simeq \frac{\kappa}{3}(1-\epsilon) a^2\bar{\rho}_{c} ~,
\ee
and solving Eq.~\eqref{Eq:FriedmannMDE2}, the Hubble rate is found to be
\be\label{Eq:HubbleSolMDE}
\mathcal{H}=\left(\frac{1-\epsilon}{1+2\epsilon}\right)\frac{2}{\eta}~.
\ee

With the above background solution, let us now turn to the perturbations.  Working in the longitudinal gauge, Eqs.~\eqref{Eq:MatterPerturbationsDeltaTheta} give
\be\label{Eq:MatterContinuityMDE}
(1-\epsilon)\delta^{\prime}_c+\theta_c-3\psi^{\prime}=0~,\hspace{1em}
\theta^{\prime}_c+\left(\frac{1-4\epsilon}{1-\epsilon}\right)\mathcal{H}\theta_c-k^2\phi-\epsilon\, k^2\delta_c=0~.
\ee
Note that in Eqs.~\eqref{Eq:MatterPerturbationsDeltaTheta}, the equation of state for dark matter is $w_c=0$ (while the effective equation of state $w_{c, {\rm eff}} \neq 0$ that arises due to diffusion is not relevant in this context).

Disregarding the contributions of baryons and relativistic matter species, from the perturbed field equations~\eqref{Eq:PerturbedEFEall} we obtain the following second order equation for $\psi$
\be\label{Eq:SecondOrderPsiMDE}
\psi^{\prime\prime}+3\mathcal{H}(1+c_{s,{\rm eff}}^2)\psi^{\prime} +\Big(2\mathcal{H}^{\prime}+\mathcal{H}^2(1+3 c_{s,{\rm eff}}^2)\Big)\psi+c_{s,{\rm eff}}^2 k^2\psi=0~,
\ee
where
\be
c_{s,{\rm eff}}^2=\frac{\epsilon}{1-\epsilon}~
\ee
is the effective speed of sound of interacting cold dark matter. We also have $\phi=\psi$, since there is no anisotropic stress. We observe that $c_{s,{\rm eff}}^2=w_{c,{\rm eff}}$~, and therefore this describes the propagation of adiabatic pressure perturbations. This is in spite of the fact that departures from the ordinary dust equation of state $w_c=0$ originate from the violation of energy conservation, as anticipated in Sec.~\ref{Sec:TransferModel}. In order to avoid gradient instabilities we must require $c_{s,{\rm eff}}^2\geq0$, that is $0\leq \epsilon<1$. Recalling the definition of $\epsilon$ given in Sec.~\ref{Sec:TransferModel}, $\epsilon\geq0$ corresponds to energy flowing from dark matter to dark energy (and, consequently, to an increasing effective cosmological constant).

Using Eq.~\eqref{Eq:HubbleSolMDE}, we can simplify Eq.~\eqref{Eq:SecondOrderPsiMDE} to
\be\label{Eq:SecondOrderPsiMDE_simplified}
\psi^{\prime\prime}+\frac{6(1+c_{s,{\rm eff}}^2)}{(1+3 c_{s,{\rm eff}}^2)\eta}\psi^{\prime} +c_{s,{\rm eff}}^2 k^2\psi=0~,
\ee
while combining Eqs.~\eqref{Eq:MatterContinuityMDE} and using $\phi=\psi$ gives
\be\label{Eq:SecondOrderMDEy}
y^{\prime\prime}+(1-3c_{s,{\rm eff}}^2)\mathcal{H}\, y^{\prime}+c_{s,{\rm eff}}^2 k^2 y+k^2(1+3c_{s,{\rm eff}}^2)\psi=0~,
\ee
having defined
\be\label{Eq:yDef}
y\coloneqq (1-\epsilon)\delta_c-3\psi~.
\ee
The second-order differential equations~\eqref{Eq:SecondOrderPsiMDE_simplified} and \eqref{Eq:SecondOrderMDEy} close a system that describes the (coupled) dynamics of the gravitational potential and density perturbations on all scales.

\subsection{Modes outside the sound horizon}

For modes on scales much larger than the sound horizon $c_{s,{\rm eff}} k\eta \ll1$, the non-decaying solution for the potential is a constant mode $\psi=\psi_0$. Equation~\eqref{Eq:SecondOrderMDEy} can then be approximated (neglecting the effective pressure term) as
\be
y^{\prime\prime}+\left(\frac{1-3c_{s,{\rm eff}}^2}{1+3c_{s,{\rm eff}}^2}\right)\frac{2}{\eta}\, y^{\prime}+k^2 \psi_0\simeq0~,
\ee
which admits a growing solution. Thus, disregarding the decaying and constant solutions of the homogeneous equation,
\be
y\simeq-\frac{1}{6}\left(\frac{1+3c_{s,{\rm eff}}^2}{1-c_{s,{\rm eff}}^2}\right)(k\eta)^2 \psi_0~.
\ee
Since $y\simeq (1-\epsilon)\delta_c$ and $a\sim\eta^{2/(1+3c_{s,{\rm eff}}^2)}$, the density contrast grows as
\be\label{Eq:DensityContrast_PowerLaw}
\delta_c \sim a^{1+3 c_{s,{\rm eff}}^2} ~,
\ee
like the generalized dark matter model of Ref.~\cite{Hu:1998kj}.  Note that the $\epsilon\to0$ limit is continuous for super-horizon modes, and gives the standard $\Lambda$CDM result.

\subsection{Modes inside the sound horizon}

For modes inside the sound horizon  $c_{s,{\rm eff}} k\eta \gg1$,  we can neglect the Hubble friction term in Eqs.~\eqref{Eq:SecondOrderPsiMDE_simplified} and \eqref{Eq:SecondOrderMDEy}, which gives
\be\label{Eq:SubHorizonMDE}
\psi^{\prime\prime}+c_{s,{\rm eff}}^2 k^2\psi\simeq0~, \hspace{1em}
y^{\prime\prime}+c_{s,{\rm eff}}^2 k^2 y+k^2(1+3c_{s,{\rm eff}}^2)\psi\simeq 0~.
\ee
The solutions for this system are
\begin{subequations}
\begin{align}
\psi=& A_k\cos( c_{s,{\rm eff}} k\eta +\varphi_k)~,\\
y= & B_k \cos( c_{s,{\rm eff}} k\eta +\tilde{\varphi}_k) -\frac{ 1+3 c_{s,{\rm eff}}^2 }{4 c_{s,{\rm eff}}^2}A_k\Big(\cos (c_{s,{\rm eff}} k
   \eta+\varphi_k )+2 (c_{s,{\rm eff}} k \eta) \sin (c_{s,{\rm eff}} k \eta+\varphi_k ) \Big) ~,
\end{align}
\end{subequations}
where $A_k$, $B_k$, $\varphi_k$, $\tilde{\varphi}_k$ depend on the initial conditions.  Keeping only the growing term in the solution for $y$,
\be
y\simeq -A_k\left(\frac{ 1+3 c_{s,{\rm eff}}^2 }{2 c_{s,{\rm eff}}^2}\right) (c_{s,{\rm eff}} k \eta) \sin (c_{s,{\rm eff}} k \eta+\varphi_k )~.
\ee
Since the potential $\psi$ does not contain any growing terms, $\delta_c\simeq (1+c_{s,{\rm eff}}^2)y$ so
\begin{align} \label{Eq:DensityContrastDM_SubHor}
\delta_c & \simeq -A_k\left(\frac{ (1+3 c_{s,{\rm eff}}^2)(1+c_{s,{\rm eff}}^2) }{2 c_{s,{\rm eff}}^2}\right) (c_{s,{\rm eff}} k \eta) \sin (c_{s,{\rm eff}} k \eta+\varphi_k ) \nonumber \\
&\simeq  -\frac{k}{c_{s,{\rm eff}}}A_k\, a^{\tfrac{1}{2}(1+3 c_{s,{\rm eff}}^2)}\sin (c_{s,{\rm eff}} k \eta+\varphi_k )~.
\end{align}
Thus, on these scales the density contrast departs from the power-law behaviour of Eq.~\eqref{Eq:DensityContrast_PowerLaw} and oscillates with increasing amplitude. For small $\epsilon$, we have $c_{s,{\rm eff}}^2\simeq \epsilon$ and Eq.~\eqref{Eq:DensityContrastDM_SubHor} shows that the amplitude of $\delta_c$ grows as%
\footnote{Note that the sine function cannot be Taylor-expanded since for sub-horizon modes $c_{s,{\rm eff}} k \eta\gg 1$.}
$\epsilon^{-1/2}$.

This behaviour should be compared with the solution for the case where one has exactly $\epsilon=0$, i.e., zero sound speed. In this case the evolution equations \eqref{Eq:SecondOrderPsiMDE_simplified}, \eqref{Eq:SecondOrderMDEy} reduce to
\be
\psi^{\prime\prime}+\frac{6}{\eta}\psi^{\prime}=0 ~, \hspace{1em}
y^{\prime\prime}+\frac{2}{\eta}\, y^{\prime}+k^2\psi=0~,
\ee
whose solutions, disregarding decaying modes, are
\be
\psi=\psi_0 ~, \hspace{1em} y\simeq -\frac{1}{6}(k\eta)^2\psi_0~.
\ee
In turn, the density contrast $\delta_c\simeq y$ evolves in this case as
\be\label{Eq:DensityScaling_ZeroEps}
\delta_c \sim a ~.
\ee
on all scales, both inside and outside the sound horizon. Comparing \eqref{Eq:DensityContrastDM_SubHor} with \eqref{Eq:DensityScaling_ZeroEps}, we conclude that the $\epsilon\to0$ limit is a singular one for perturbations inside the sound horizon.
  
Note that for any fixed $c_{s,{\rm eff}}$, the ratio of the amplitudes in \eqref{Eq:DensityScaling_ZeroEps} to \eqref{Eq:DensityContrastDM_SubHor} is monotonically increasing.  This shows that, given equal initial amplitudes, structure formation will occur more rapidly in the standard $\Lambda$CDM cosmology---this is to be expected since in $\Lambda$CDM there is no pressure from cold dark matter to counteract gravitational attraction, while in this model the interacting dark matter has a small effective pressure $p_{c, {\rm eff}} \simeq \epsilon\, \rho_c$.

In our model, cold dark matter effectively behaves as in the phenomenological model for generalized dark matter \cite{Hu:1998kj} in the case of zero viscosity; recent studies of the detailed properties of this model and constraints on its parameters can be found in Refs.~\cite{Thomas:2016iav,Kopp:2016mhm,Ilic:2020onu}. The upper bounds on $c_{s,{\rm eff}}^2$ obtained in Ref.~\cite{Thomas:2016iav} gives the constraint
\be
\epsilon ~ \lesssim ~ \mathcal{O}(10^{-6}).
\ee

\section{Other transfer models}
\label{Sec:Generalizations}

Our analysis of the evolution of cosmological perturbations is based on a simple model for the covariant energy-momentum transfer introduced in Sec.~\ref{Sec:TransferModel}. Nevertheless, the framework developed here is fully general and can accommodate different models for the energy-momentum transfer potential $Q$ in Eq.~\eqref{Eq:DefQ}. For instance, a similar analysis could be done for
\be\label{Eq:Model2}
J_a=- \epsilon\, \alpha \nabla_a \theta_{\rm\scr CDM}~, 
\ee
where the transfer potential is proportional to the expansion scalar $\theta_{\rm\scr CDM} = \nabla_a u^a_{\rm\scr CDM}$ of cold dark matter geodesics and $\alpha$ is a coupling with dimensions of inverse length, or for the model
\be\label{Eq:Model2bis}
J_a=- \epsilon\, \nabla_a (\theta_{\rm\scr CDM})^2~, 
\ee
which does not require a dimensionful coupling.  (Although at the background level $\theta_{\rm\scr CDM}$ is three times the Hubble rate, this is not true when including perturbations.)

Generalizations combining features of models \eqref{Eq:DiffusionModelJ} and \eqref{Eq:Model2} are also possible
\be\label{Eq:Model3}
J_a=- \epsilon\, \kappa\, \nabla_a \Big(F(T_{uu}^{\rm\scr CDM}, \theta_{\rm\scr CDM}) \Big)~,
\ee
where $F$ is some scalar function that depends on both the co-moving dark matter energy density and the expansion scalar for the geodesics of cold dark matter particles. In addition to an effective equation of state $w_{\rm eff}\neq 0$, such a model may lead to cold dark matter effectively behaving as an imperfect fluid.

Further extensions are also possible, for instance by adding interactions between dark energy and other matter fields than dark matter (say radiation, baryonic matter, or an inflaton field), and/or by allowing for the transfer potential $Q$ to depend on other geometric invariants of matter world-line congruences, or space-time curvature invariants.

As seen for the model \eqref{Eq:DiffusionModelJ}, these generalizations are also expected to give rise to further non-trivial features in the effective fluid description of dark matter in addition to a non-zero adiabatic sound speed. The precise characterization of the set of transfer models that are both fully consistent with the gravitational field equations and free from instabilities is left for future work.

\section{Equivalence with generalized dark matter models}
\label{Sec:ConnectionGDM}

We have already observed in previous sections that the transfer model \eqref{Eq:DiffusionModelJ} implies that dark matter effectively behaves as in the generalized dark matter model \cite{Hu:1998kj} in the inviscid case. From a more general perspective, it is therefore natural to ask whether a similar correspondence also exists beyond linear perturbations and for more general choices of transfer potential. We will show in this section that the answer to both questions is affirmative.

The description of a mixture of two interacting adiabatic fluid with generic equations of state as a generalized dark matter model has been previously given in Ref.~\cite{Kopp:2016mhm}, which considered some specific models for the energy-momentum transfer. Here we focus specifically on the case of interacting dark energy with $w_x=-1$.

For our purposes, it is sufficient to assume the Einstein field equations $G_{ab}=\kappa\, T_{ab}$ of general relativity as a starting point, with total stress-energy tensor including the contributions of ordinary matter, dark matter, and a dark energy fluid with $w_x=-1$,
\be\label{Eq:Total_Tab}
T_{ab}=T^{\rm m}_{ab}+T^{\scr\rm DM}_{ab}+T^{\scr\rm DE}_{ab}~,~~\mbox{with}~~ T^{\scr\rm DE}_{ab}=-\rho_x \, g_{ab}~.
\ee
For simplicity, we will assume that only dark matter interacts with dark energy, while the stress-energy of ordinary matter fields is conserved separately $\nabla^c T^{\rm m}_{ac}=0$\,.
The Bianchi identies then give $\nabla^c T^{\scr\rm DM}_{ac}=\nabla_a \rho_x$\,, so the energy-momentum transfer is necessarily integrable.

Assuming that $\rho_x$ has a finite limit at future infinity (this is reasonable to expect since $\rho_x$ is dynamical only due to interactions with dark matter, and dark matter dilutes to a vanishing energy density in the distant future of an expanding universe), we can split it into two terms
\be
\rho_x=\tilde{\rho}_x-\kappa^{-1} \Lambda_{\rm f}~,
\ee
where $\tilde{\rho}_x$ tends to zero in the far future. This allows us to split dark energy into a dynamical portion $\tilde \rho_x$ and a constant contribution $-\kappa^{-1} \Lambda_{\rm f}$ (for instance, for the model~\eqref{Eq:DiffusionModelJ} studied in previous sections $\tilde{\rho}_x=-\epsilon\, \rho_c$).
The most general form for the stress-energy tensor of dark matter as a fluid is
\be
T^{\scr\rm DM}_{ab}=(\rho_d+p_d)u_a u_b + 2\, q_{(a} u_{b)} + p_d\, g_{ab}+\pi_{ab}~,
\ee
we use the index $d$ to denote dark matter that (could be but) is not necessarily cold dark matter. As in Sec.~\ref{Sec:Framework} we define the four-velocity $u_a$ so the momentum flux $q_a$ relative to $u_a$ vanishes, and then the total stress-energy tensor of the interacting dark sector can be rewritten as
\be\label{Eq:ReintepretIDE}
T^{\scr\rm DM}_{ab}+T^{\scr\rm DE}_{ab}=\check{T}_{ab}-\kappa^{-1} \Lambda_{\rm f}\, g_{ab}~,
\ee
where
\be
\check{T}_{ab}=(\check{\rho}+\check{p})u_a u_b + \check{p}\, g_{ab}+\pi_{ab}~,
\ee
with
\be\label{Eq:RhoP_redef}
\check{\rho}=\rho_d + \tilde{\rho}_x~, \quad  \check{p}=p_d-\tilde{\rho}_x~.
\ee
By construction, $\nabla^b \check{T}_{ab}=0$. Equation~\eqref{Eq:ReintepretIDE} shows that IDE models with $w_x=-1$ are mathematically equivalent to generalized dark matter models: the contributions of dynamical dark energy can be re-interpreted as corrections to the stress-energy of generalized dark matter and a cosmological constant. The effective stress-energy tensor $\check{T}_{ab}$ describes a fluid with the effective equation of state
\be
\check{w}=\frac{\check{p}}{\check{\rho}}=\frac{p_d-\tilde{\rho}_x}{\rho_d +\tilde{\rho}_x}~.
\ee
For simplicity, we assume that $T^{\scr\rm DM}_{ab}$ originally describes an adiabatic fluid.

In general, $\tilde{\rho}_x$ may depend on the physical properties of ordinary matter as well as on extra fields (including geometric observables), here denoted as $X$. We will therefore assume for generality $\tilde{\rho}_x=\tilde{\rho}_x(\rho_d,X)$~. In this more general case, the continuity equations for the effective dark matter fluid need to be supplemented by the equations of motion for the fields~$X$, which must be consistent with the gravitational field equations. The adiabatic sound speed of the effective fluid is
\be
\check{c}_a^2=\frac{\check{p}^{\prime}}{\check{\rho}^{\prime}}=\frac{\left(c_{a,d}^2-\frac{\pa\tilde{\rho}_x}{\pa \rho_d}\right)\rho_d^{\prime}-\frac{\pa\tilde{\rho}_x}{\pa X} X^{\prime}}{\left(1+\frac{\pa\tilde{\rho}_x}{\pa \rho_d}\right)\rho_d^{\prime}+\frac{\pa\tilde{\rho}_x}{\pa X} X^{\prime}}~,
\ee
while the gauge-invariant entropy perturbation is
\be
\mathcal{S}=\mathcal{H}\left(\frac{\delta \check{p}}{\check{p}^{\prime}}-\frac{\delta \check{\rho}}{\check{\rho}^{\prime}} \right)=\frac{\mathcal{H}}{\check{p}^{\prime}} \left[\left(\frac{c_{a,d}^2-\frac{\pa\tilde{\rho}_x}{\pa \rho_d} }{1+\frac{\pa\tilde{\rho}_x}{\pa \rho_d}} -\check{c}_a^2 \right)\delta \check{\rho}-\left(\frac{1+c_{a,d}^2}{1+\frac{\pa\tilde{\rho}_x}{\pa \rho_d}}\right)\frac{\pa\tilde{\rho}_x}{\pa X}\,\delta X   \right]~.
\ee
Note that, even if we assume that $T^{\scr\rm DM}_{ab}$ originally describes an adiabatic fluid, in general $\mathcal{S}\neq0$ unless $X$ evolves adiabatically with respect to $\rho_d$. The sound speed $\check{c}_s^2$ and the intrinsic pressure perturbation $\mathcal{S}$ are related as follows
\be
\mathcal{S}=\left(1-\frac{\check{c}_s^2}{\check{c}_a^2}\right)\left[\frac{\check{\delta}}{3(1+\check{w})}+k^{-2} \mathcal{H} \check{\theta}\right]~.
\ee
Note that shear viscosity only arises if $\pi_{ab} \neq 0$ of the original dark matter fluid; it is unaffected by the energy-momentum transfer with dark energy.

In the special case where $\tilde{\rho}_x=\tilde{\rho}_x(\rho_d)$, the intrinsic entropy perturbation of the effective fluid vanishes and
\be
\check{c}_s^2=\check{c}_a^2=\frac{c_{a,d}^2-\frac{\pa\tilde{\rho}_x}{\pa \rho_d}}{1+\frac{\pa\tilde{\rho}_x}{\pa \rho_d}} ~, \quad \check{w}=\frac{w \rho_d -\tilde{\rho}_x(\rho_d)}{\rho_d +\tilde{\rho}_x(\rho_d)}~,
\ee
both of which are in general time dependent. In this special case, it is sufficient to assign the functional form of $\tilde{\rho}_x(\rho_d)$ in order to have a fully consistent non-perturbative definition of a generalized dark matter model.

As seen in Sec.~\ref{Sec:GMDE}, the IDE model proposed in Sec.~\ref{Sec:TransferModel} can be rewritten as a generalized dark matter model with $\check{w} = \check{c}_s^2 = \check{c}_a^2 = \epsilon/(1-\epsilon)$.  Another example is the IDE model with background transfer $\kappa^{-1} J_0=\rho_x^{\,\prime}=-\alpha \mathcal{H} \rho_c$ (omitting overbars) between dark energy and pressureless cold dark matter.  Using $\rho_c^{\prime}+3\mathcal{H}\rho_c=0$ to compute $\rho_c^{\prime}$, this second example is equivalent to a generalized dark matter model with $\check{\rho}=(1+\alpha/3)\rho_c$, $\check{p}=- \alpha \rho_c / 3$, and $\check{w} = -\alpha/(\alpha + 3)$ (while $\check{c}_s^2$ and $\check{c}_a^2$ depend on how $\rho_x^{\,\prime}=-\alpha \mathcal{H} \rho_c$ is extended to include perturbations).

One consequence of this equivalence between certain IDE and generalized dark matter models is that we may use observational constraints derived for some generalized dark matter models and translate them into constraints for IDE models with $w_x=-1$, and vice versa. These may in turn be used to constrain the functional form of $\tilde{\rho}_x$. This equivalence also offers a direct procedure to build generalized dark matter models starting from an IDE. We leave a detailed analysis of generalized dark matter models built from this perspective for future work.

\section{Summary}
\label{Sec:Conclusions}

In this work, we carried out a detailed analysis of scalar cosmological perturbations in unimodular gravity (based on the trace-free Einstein field equations), without assuming stress-energy conservation and without imposing any further constraints, such as the unimodularity condition. In this theory, the gravitational field equations are equivalent to interacting dark energy models with integrable energy-momentum transfer and a dark-energy equation of state $w_x=-1$. The formalism developed in Sec.~\ref{Sec:Framework} does not rely on any further assumptions on the specific form of the energy-momentum transfer. Therefore, it applies to all IDE models with $w_x=-1$ (for both the trace-free Einstein equations as well as for the standard Einstein equations of general relativity).

We then considered a specific model in Sec.~\ref{Sec:TransferModel}, where the energy-momentum transfer potential is proportional to the energy density of cold dark matter (and dark matter is assumed to be the only matter species responsible for energy-momentum non-conservation). The ensuing dark-sector interactions generate a non-zero effective sound speed $c_{s,{\rm eff}}$ for the dark matter.  We studied the evolution of perturbation modes during the radiation and matter dominated eras, and we find that in this model (i) super-horizon perturbations avoid the instability found in Ref.~\cite{Valiviita:2008iv} due to the absence of non-adiabatic pressure perturbation, as explained in Sec.~\ref{Sec:Dsummary}, (ii) the energy transfer must be from dark matter to dark energy in order to avoid a gradient instability, (iii) there are small departures from the dynamics of standard $\Lambda$CDM cosmology that are parametrized by the amplitude of energy-momentum non-conservation, and (iv) for modes inside the sound horizon during matter-domination the $c_{s,{\rm eff}} \to 0$ limit is singular, making this a particularly interesting regime to explore further the differences between this model and the standard $\Lambda$CDM cosmology. Further, in this model cold dark matter effectively behaves as in the phenomenological model for generalized dark matter \cite{Hu:1998kj} (in the special case of zero viscosity), which has been examined in detail in Refs.~\cite{Thomas:2016iav,Kopp:2016mhm,Ilic:2020onu}. The constraints obtained in Ref.~\cite{Thomas:2016iav} give an upper bound $\epsilon\lesssim \mathcal{O}(10^{-6})$ on the dimensionless coupling constant for the transfer model \eqref{Eq:DiffusionModelJ}.

In Sec.~\ref{Sec:Generalizations} we describe other possible IDE models that could also be studied with the same formalism. Finally, we showed in Sec.~\ref{Sec:ConnectionGDM} that the family of interacting dark energy models considered in this paper are equivalent to a class of generalized dark matter models. This equivalence offers the opportunity to translate constraints derived for generalized dark matter into constraints for IDE models with $w_x=-1$, and vice versa.

\section*{Acknowledgments}

We thank Roy Maartens for comments on an earlier draft of this paper.
The work of MdC was supported under grants No.~FIS2017-85076-P (Spanish MINECO/AEI/FEDER, UE) and No.~IT956-16 (Basque Government), and by the Natural Sciences and Engineering Research Council of Canada and the Atlantic Association for Research in the Mathematical Sciences (AARMS).  The work of EWE was supported in part by the Natural Sciences and Engineering Research Council of Canada, and the UNB Fritz Grein Research Award.

\bibliography{unimodular_references}

\end{document}